\documentclass[letterpaper]{article} % DO NOT CHANGE THIS
\usepackage{aaai24}  % DO NOT CHANGE THIS
\usepackage{times}  % DO NOT CHANGE THIS
\usepackage{helvet}  % DO NOT CHANGE THIS
\usepackage{courier}  % DO NOT CHANGE THIS
\usepackage[hyphens]{url}  % DO NOT CHANGE THIS
\usepackage{graphicx} % DO NOT CHANGE THIS
\usepackage{natbib}  % DO NOT CHANGE THIS AND DO NOT ADD ANY OPTIONS TO IT
\usepackage{caption} % DO NOT CHANGE THIS AND DO NOT ADD ANY OPTIONS TO IT
\usepackage{algorithm}
\usepackage{algorithmic}
\usepackage{amsmath}
\usepackage{amsfonts}
\usepackage{color}
\usepackage{booktabs}
\usepackage{caption}
\usepackage{subcaption}

\usepackage{newfloat}
\usepackage{listings}
\DeclareCaptionStyle{ruled}{labelfont=normalfont,labelsep=colon,strut=off} % DO NOT CHANGE THIS
\floatstyle{ruled}
\newfloat{listing}{tb}{lst}{}
\floatname{listing}{Listing}
\usepackage{bm} % added by Y. Takano
\title{Robust portfolio optimization model for electronic coupon allocation}
\author {
    Yuki Uehara\textsuperscript{\rm 1},
    Naoki Nishimura\textsuperscript{\rm 1,2},
    Yilin Li\textsuperscript{\rm 3},
    Jie Yang\textsuperscript{\rm 3},
    Deddy Jobson\textsuperscript{\rm 3},
    Koya Ohashi\textsuperscript{\rm 3},\\
    Takeshi Matsumoto\textsuperscript{\rm 3},
    Noriyoshi Sukegawa\textsuperscript{\rm 4},
    Yuichi Takano\textsuperscript{\rm 1}
}
\affiliations {
    \textsuperscript{\rm 1}University of Tsukuba\\
    \textsuperscript{\rm 2}Recruit Co., Ltd\\
    \textsuperscript{\rm 3}Mercari, inc.\\
    \textsuperscript{\rm 4}Hosei University\\
    s2320413@u.tsukuba.ac.jp, 
    nishimura@r.recruit.co.jp,\\
    \{y-li, j-yang, deddy, k-ohashi, takeshi.matsumoto\}@mercari.com,\\
    sukegawa@hosei.ac.jp,
    ytakano@sk.tsukuba.ac.jp
}

\usepackage{bibentry}
\begin{document}

\maketitle

\begin{abstract}
Currently, many e-commerce websites issue online/electronic coupons as an effective tool for promoting sales of various products and services. 
We focus on the problem of optimally allocating coupons to customers subject to a budget constraint on an e-commerce website. 
We apply a robust portfolio optimization model based on customer segmentation to the coupon allocation problem. 
We also validate the efficacy of our method through numerical experiments using actual data from randomly distributed coupons. 
Main contributions of our research are twofold. 
First, we handle six type of coupons, thereby making it extremely difficult to accurately estimate the difference in the effects of various coupons.
Second, we demonstrate from detailed numerical results that the robust optimization model achieved larger uplifts of sales than did 
%\color{blue}
the commonly-used multiple-choice knapsack model and 
%\color{black}
the conventional mean--variance optimization model. 
Our results open up great potential for robust portfolio optimization as an effective tool for practical coupon allocation. 
\end{abstract}

\section{Introduction}
\label{intro}

\subsection{Background}

In the e-commerce marketplace, marketers implement various types of treatments (or sales promotions) as a part of the customer relationship management.
Currently, many e-commerce websites issue online/electronic coupons as an effective tool for promoting sales of various products and services.
These coupons enable personalized/differentiated pricing and thus benefit both consumers and businesses~\cite{sigala2013framework}.
For consumers, coupons promise substantial cost savings.
For businesses, coupons help boost demand, increase sales, promote brand awareness, and acquire new customers. 
In fact, the impact of coupons on sales has been demonstrated in the literature (e.g., \citet{duan2022online}). 

Business analytics aims to enable organizations to make smart decisions for creating business value through extensive data utilization.
In particular, prescriptive analytics, which seeks to find the best course of action for the future, is receiving plenty of attention from the machine learning and operations research communities~\cite{bertsimas2020predictive,lepenioti2020prescriptive,lo2023meaningful}. 
For example, this framework helps in price optimization, combined with machine learning methods for demand forecasting~\cite{ikeda2023prescriptive,ito2017optimization}. 
We focus on applying prescriptive analytics to the problem of optimally allocating coupons to customers on an e-commerce website.

\subsection{Related work}
\citet{li2020spending} formulated the budget-constrained coupon allocation problem as a multiple-choice knapsack problem~\cite{kellerer2004multiple} using deep neural networks for user intent detection. 
Several recent studies also approached multiple-choice knapsack problems for personalized promotions using uplift modelling estimations~\cite{albert2022commerce}, direct and enduring effect predictions~\cite{yang2022personalized}, and causal forest algorithms~\cite{ai2022lbcf}. 
However, since the knapsack problem is known to be NP-hard, it is clearly difficult to solve practical large-scale problems efficiently. 
%\citet{makhijani2019lore}

The coupon allocation can be regarded as a special case of marketing budget allocation~\cite{zhao2019unified}. 
Therefore, \citet{lo2015predictive} introduced the mean--variance portfolio optimization framework~\cite{markowitz1959portfolio} for allocating marketing campaigns into customer segments. 
In addition, \citet{pachamanova2020uncertainty} extended this portfolio optimization framework to robust optimization~\cite{ben2009robust,bertsimas2011theory} under uncertainty about estimated treatment effects. 
One of the greatest advantages of this framework is the computational load that is significantly reduced by converting large-sized combinatorial optimization models (i.e., multiple-choice knapsack problems) to small-sized continuous portfolio optimization models. 

However, the effectiveness of this portfolio optimization framework remains undemonstrated and has never been applied to allocation of multiple types of coupons. 
Specifically, \citet{lo2015predictive} did not evaluate the uplift performance of portfolio optimization models, and \citet{pachamanova2020uncertainty} failed to show that the robust optimization model could improve the average uplift performance. 
Additionally, these prior studies dealt with only two types of treatments (i.e., e-mails of men's and women's merchandise). 

\subsection{Our contribution}
A primary goal of this paper is to focus spotlight on the application of the portfolio optimization framework to coupon allocation. 
Specifically, we apply a robust portfolio optimization model to the coupon allocation problem and validate the efficacy of our method through numerical experiments. 
We evaluate the uplift performance of our method using actual data of randomly distributed coupons obtained from 616,000 customers on an online marketplace app. 

Main contributions of our research are twofold. 
First, we handle six type of coupons (i.e., 5\% discount, 5\% cashback, 10\% discount, 10\% cashback, 15\% discount, and 300 yen discount). 
In this case, it is extremely difficult to accurately estimate the difference in the effects of various coupons.
Second, we demonstrate from detailed numerical results that the robust optimization model achieved larger uplifts of sales than did 
%\color{blue}
the commonly-used multiple-choice knapsack model and 
%\color{black}
the conventional mean--variance optimization model. 
Our results open up great potential for robust portfolio optimization as an effective tool for practical coupon allocation. 

\section{Optimization models}
\label{formulation}

In this section, we first explain how to estimate coupon effects and 
%\color{blue}
next introduce the multiple-choice knapsack model as a baseline for numerical experiments. 
%\color{black}
We then present two types of portfolio optimization models for coupon allocation. 

\subsection{Coupon effect estimation}
Uplift modeling is a type of predictive modeling that aims at predicting the net effect of performing some action on a certain outcome~\cite{devriendt2018literature}. 
We use the metalearnrs~\cite{kunzel2019metalearners} that combine machine learning and causal inference to estimate the conditional average treatment effect (CATE). 

For each customer, let $X$ denote a vector of predictor variables, and $Y(j)$ denote an outcome variable (e.g., purchase amount) depending on coupon $j \in J$, where no coupons are allocated when $j=0$. 
The CATE, the effect of allocating coupon $j$ to a customer with the data $X=x$ of predictor variables, is then given by 
\begin{align*}
\pi(x,j) = {\rm E}[Y(j)-Y(0) \mid X=x].
\end{align*}

The S-learner, which a typical algorithm of the metalearnrs~\cite{kunzel2019metalearners}, constructs a machine learning model to estimate ${\rm E}[Y(j) \mid X=x]$ for each coupon $j \in \{0\} \cup J$ and then uses 
\begin{align*}
\hat{\pi}(x,j) = {\rm E}[Y(j) \mid X=x] - {\rm E}[Y(0) \mid X=x]
\end{align*}
as an estimator of the CATE. 

%\color{blue}
\subsection{Multiple-choice knapsack (MCK) model}
Let $\bm{v}=(v_{ij})_{(i,j) \in I \times J} \in \mathbb{R}^{|I \times J|}$ be a binary design variable for allocating coupons $j \in J$ to customers $i \in I$, where $v_{ij} = 1$ if type $j$ coupon is allocated to customer $i$, and $v_{ij} = 0$ otherwise. 
Let $B$ denote a total budget for coupon allocation and $c_j$ denote a unit cost of coupon $j \in J$.

In the multiple-choice knapsack (MCK) model, the total coupon effect is maximized subject to the constraints on total budget and one-customer-one-coupon allocation.  
When $\hat{\pi}_{ij} = \hat{\pi}(x_i,j)$ is an estimated effect of coupon $j \in J$ on customer $i \in I$ with $X = x_i$, the MCK model for coupon allocation~\cite{li2020spending} is formulated as follows: 
%This model uses a $0$-$1$ variable $v_{ij}$ taking one if customer $i \in I$ receives coupon $j$.
%The design variable is thus $\bm{v}=(v_{ij})_{(i,j) \in I \times J} \in \{0,1\}^{|I \times J|}$ and the formulation is given as: 
\begin{equation*}
\begin{aligned}
&\max_{\bm{v}}	&& 
\sum_{i \in I}\sum_{j \in J} \hat{\pi}_{ij} v_{ij}\\
&{\rm ~s.~t.}	&&\sum_{i\in I}\sum_{j \in J} c_{j} v_{ij} \le B,\\
&				&&\sum_{j \in J} v_{ij} \le 1\quad (i \in I),\\
&				&&v_{ij} \in \{0,1\}\quad (i \in I, j \in J).
\end{aligned}
\end{equation*}
%where $\bar{\pi}_{ij}$ is the value of $\hat{\pi}(x,j)$ with $x$ being the vector of predictor variables for user $i$. 
%\color{black}

\subsection{Mean--variance optimization (MVO) model}

According to the portfolio optimization framework~\cite{lo2015predictive}, we perform customer segmentation based on cluster analysis. 
As a result of the cluster analysis, we denote by $N_k$ the number of customers belonging to cluster $k \in K$. 
By regarding these clusters as targets of investment, portfolio optimization models can be naturally applied to coupon allocation. 

Let $\bm{w}=(w_{kj})_{(k,j) \in K \times J} \in \mathbb{R}^{|K \times J|}$ be a continuous design variable for coupon allocation, where $w_{kj}$ is the number of type $j$ coupons to be allocated to cluster $k$. 
When $\pi_{kj}$ is an effect of coupon $j \in J$ on cluster $k \in K$, the mean and variance of the total coupon effect can be estimated as follows: 
\begin{align}
\mathrm{E}[\sum_{k \in K}\sum_{j \in J} \pi_{kj} w_{kj}] & \approx \sum_{k \in K}\sum_{j \in J} \bar{\pi}_{kj} w_{kj}, \notag \\
\mathrm{Var}[\sum_{k \in K}\sum_{j \in J} \pi_{kj} w_{kj}] & \approx \sum_{k,t\in K}\sum_{j,s\in J} \sigma_{kjts} w_{kj} w_{ts}, \notag 
\end{align}
where $\bar{\pi}_{kj}$ is the sample mean of $\hat{\pi}(x,j)$ in cluster $k \in K$, and $\sigma_{kjts}$ is the covariance between $\pi_{kj}$ and $\pi_{ts}$. 
These covariances were estimated using the bootstrap procedure in \citet{lo2015predictive}. 

In the mean--variance optimization (MVO) model, the weighted sum of the mean and variance of the total coupon effect is maximized subject to the constraints on total budget and cluster sizes. 
%Let $B$ denote a total budget for coupon allocation and $c_j$ denote a unit cost of coupon $j \in J$. 
As in \citet{lo2015predictive}, the MVO model for coupon allocation can be formulated as follows: 
\begin{equation*}
\begin{aligned}
&\max_{\bm{w}}	&& (1 - \lambda) \sum_{k \in K}\sum_{j \in J} \bar{\pi}_{kj} w_{kj} - \lambda \sum_{k,t\in K}\sum_{j,s\in J} \sigma_{kjts} w_{kj} w_{ts} \\
&{\rm ~s.~t.}	&&\sum_{k\in K}\sum_{j \in J} c_{j} w_{kj} \le B,\\
&				&&\sum_{j \in J} w_{kj} \le N_k\quad (k \in K),\\
&				&&w_{kj} \ge 0\quad (k \in K, j \in J),
\end{aligned}
\end{equation*}
where $\lambda \in [0,1]$ is a hyperparameter of risk aversion. 

\subsection{Robust optimization (RO) model}
\label{ours}
It is known that the MVO model often performs badly because the associated parameters (i.e., $\bar{\pi}_{kj}$ and $\sigma_{kjts}$) are subject to estimation errors~\cite{broadie1993computing}. 
To remedy this situation, we focus on robust optimization~\cite{ben2009robust,bertsimas2011theory} for maximizing the worst-case performance under parameter uncertainty. 

We focus on the following interval uncertainty sets of coupon effects: 
\begin{align*}
\pi_{kj} \in [\bar{\pi}_{kj}-\alpha \delta_{kj}, \bar{\pi}_{kj}] \quad (k \in K, j \in J),
\end{align*}
where $\delta_{kj}$ is the standard error of the sample mean $\bar{\pi}_{kj}$, and $\alpha$ is a hyperparameter of the interval scale. 
%These standard errors can be estimated using the bootstrap procedure. 
It is, however, unlikely that the worst case (i.e., $\pi_{kj} = \bar{\pi}_{kj}-\alpha \delta_{kj}$) happens at the same time for all $(k,j) \in K \times J$. 
Therefore, \citet{bertsimas2003robust} introduced the budgeted uncertainty set specified by $\Gamma$, a hyperparameter of the cardinality to ensure that the worst case will happen for at most $\Gamma$ pairs of $(k,j) \in K \times J$.  

Let $S \subseteq K \times J$ be a subset of cluster--coupon pairs for which the worst case happens, namely, 
\[
\pi_{kj}=
\begin{cases}
\bar{\pi}_{kj}-\alpha \delta_{kj} & \mbox{if $(k,j) \in S$}, \\
\bar{\pi}_{kj} & \mbox{otherwise}. 
\end{cases}
\]
We then consider maximizing the following total coupon effect in the worst case:
\begin{align*}
\sum_{k\in K}\sum_{j \in J} \bar{\pi}_{kj} w_{kj}
+\min_{S} \{ \sum_{(k,j)\in S} -\alpha\delta_{kj} w_{kj} \mid |S|\le \Gamma \}.
\end{align*}
Note here that the inner minimization problem can be rewritten as
\begin{equation*}
\begin{aligned}
&\min_{\bm{z}}	&&\sum_{k\in K}\sum_{j \in J} -\alpha\delta_{kj} w_{kj} z_{kj}\\
&{\rm ~s.~t.}	&&\sum_{k\in K}\sum_{j \in J} z_{kj} \le \Gamma,\\
&				&& 0 \le z_{kj} \le 1 \quad (k \in K, j \in J),
\end{aligned}
\end{equation*}
where $\bm{z} = (z_{kj})_{(k,j) \in K \times J} \in \mathbb{R}^{|K \times J|}$ is an auxiliary design variable serving as a subset $S \subseteq K \times J$. 

By transforming the inner minimization problem into a dual maximization problem, we obtain the following robust optimization (RO) model for coupon allocation~\cite{bertsimas2003robust,pachamanova2020uncertainty}: 
\begin{equation*}
\begin{aligned}
&\max_{\bm{w},h,\bm{q}}	&&\sum_{k\in K}\sum_{j \in J} \bar{\pi}_{kj} w_{kj}
- \Gamma h - \sum_{k\in K}\sum_{j \in J} q_{kj}\\
&{\rm ~s.~t.}	&&\sum_{k\in K}\sum_{j \in J} c_{j} w_{kj} \le B,\\
&				&&\sum_{j \in J} w_{kj} \le N_k\quad (k \in K),\\
&				&&h+q_{kj} \ge \alpha\delta_{kj} w_{kj} \quad (k \in K, j \in J),\\
&				&&w_{kj} \ge 0\quad (k \in K, j \in J),\\
&				&&h \ge 0,\\
&				&&q_{kj} \ge 0 \quad (k \in K, j \in J),
\end{aligned}
\end{equation*}
where $h \in \mathbb{R}$ and $\bm{q}=(q_{kj})_{(k,j) \in K \times J} \in \mathbb{R}^{|K \times J|}$ are dual design variables. 

\begin{figure*}[tb]
\centering
\begin{subfigure}[t]{0.33\textwidth}
\includegraphics[width=\textwidth]
{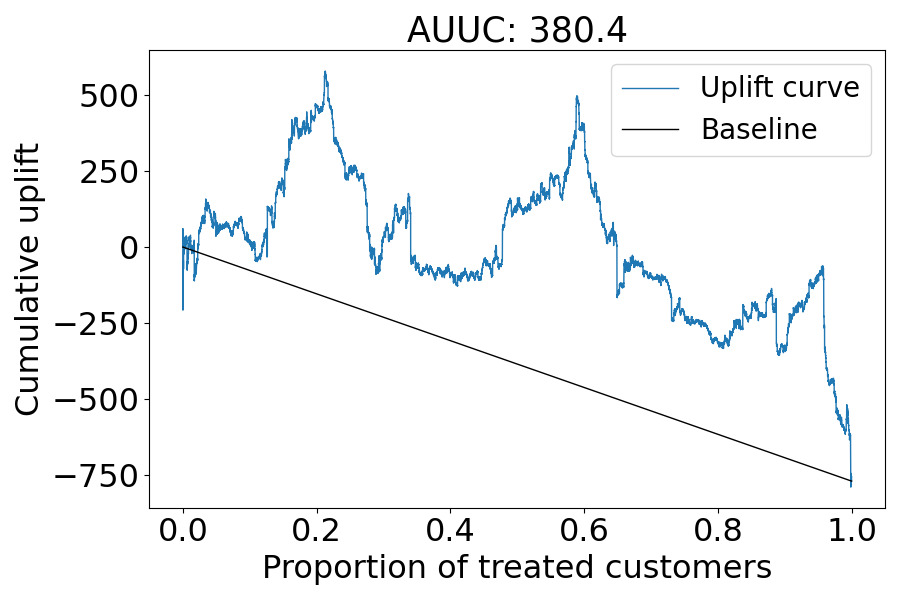}
\caption{5\% discount}
\end{subfigure}
\begin{subfigure}[t]{0.33\textwidth}
\includegraphics[width=\textwidth]
{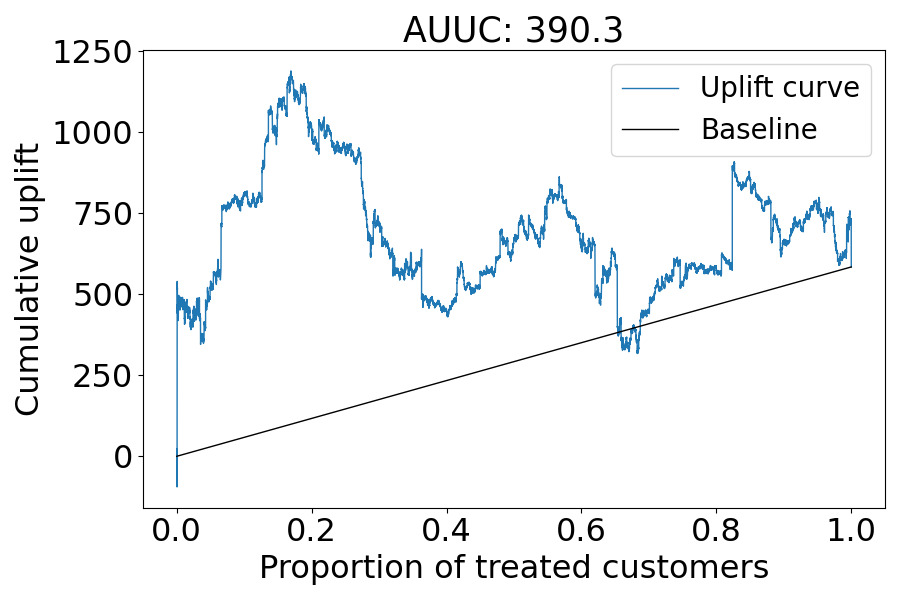}
\caption{5\% cashback}
\label{fig:figa11}
\end{subfigure}
\begin{subfigure}[t]{0.33\textwidth}
\includegraphics[width=\textwidth]
{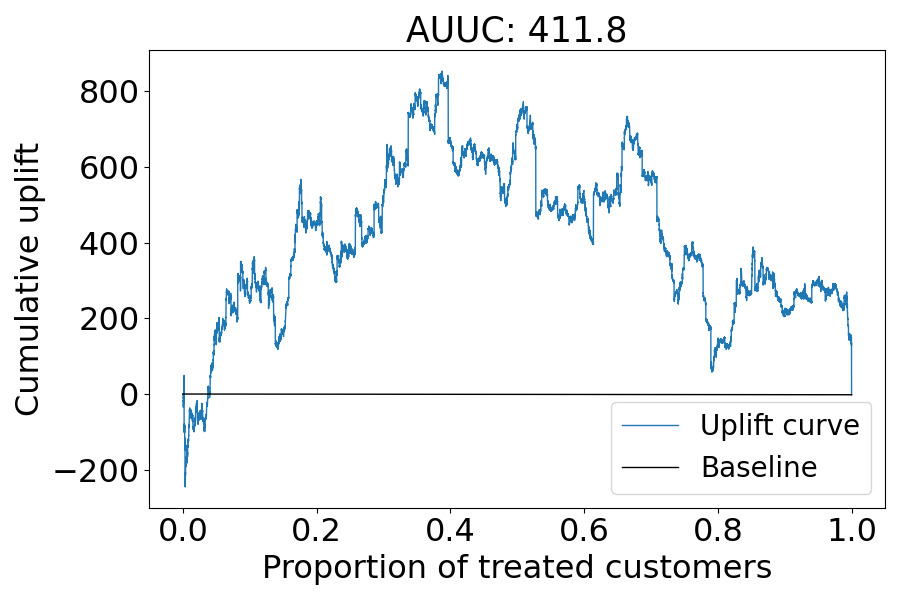}
\caption{10\% discount}
\label{fig:figa11}
\end{subfigure}\\[0.5cm]
\begin{subfigure}[t]{0.33\textwidth}
\includegraphics[width=\textwidth]
{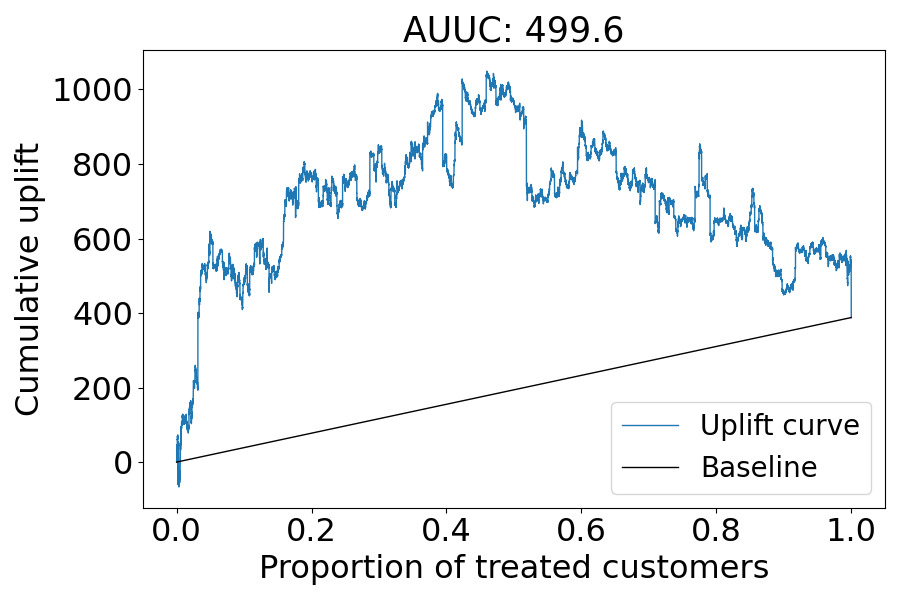}
\caption{10\% cashback}
\label{fig:figa11}
\end{subfigure}
\begin{subfigure}[t]{0.33\textwidth}
\includegraphics[width=\textwidth]
{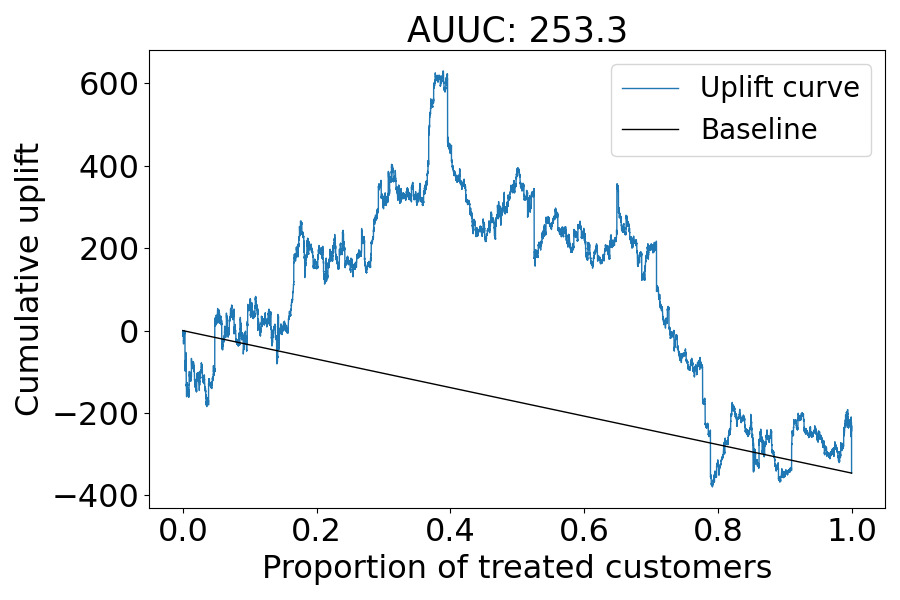}
\caption{15\% discount}
\label{fig:figa11}
\end{subfigure}
\begin{subfigure}[t]{0.33\textwidth}
\includegraphics[width=\textwidth]
{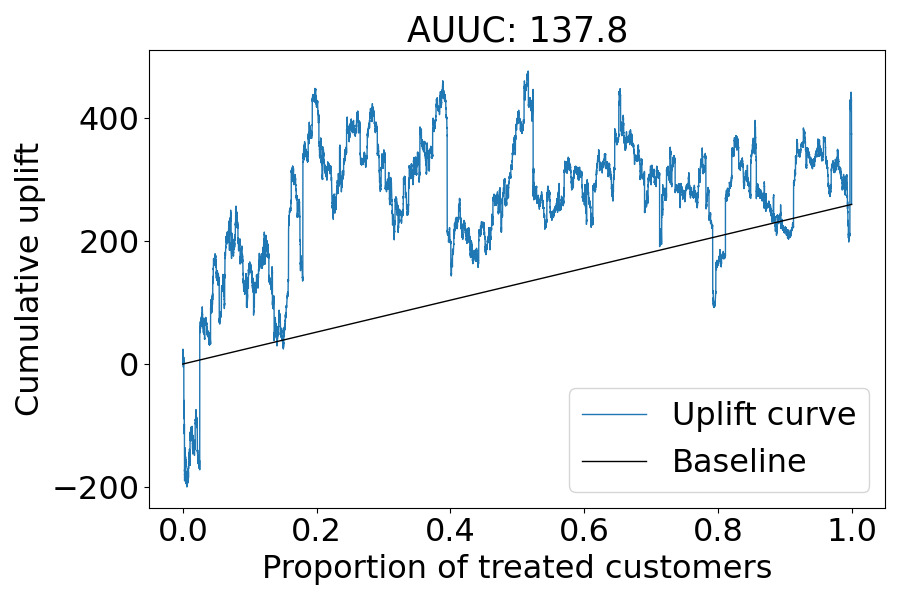}
\caption{300 yen discount}
\label{fig:figa11}
\end{subfigure}
\caption{Uplift curves of GMV for six types of coupons}
\label{fig:up}
\end{figure*}

\section{Numerical experiments}
\label{experiment}

In this section, we evaluate the effectiveness of the RO model for coupon allocation through numerical experiments.
Note that for confidentiality reasons, monetary  (e.g., sales and budget) values are normalized based on the overall average. 

\subsection{Dataset}

We used actual data from randomly distributed coupons on an online marketplace app operated by Mercari Inc., a Japanese e-commerce company.
Our dataset consists of 616,000 customers, where 528,000 customers were randomly sampled to receive one of six types of coupons (Table~\ref{tbl:coupon}) (i.e., 88,000 recipients of each type of coupons), and the remaining 88,000 customers received no coupons. 

For each customer $i \in I$, the gross merchandise volume ($\mathrm{GMV}_i$) within two weeks after coupon allocation was employed as the outcome variable, and 95 features (e.g., purchase history before coupon allocation) were used as predictor variables for coupon effect estimation. 
This dataset of 616,000 customers was randomly split into training (80\%) and testing (20\%) datasets. 

\begin{table}[h]
\centering
\caption{Coupon types and their costs}
\label{tbl:coupon}
\begin{tabular}{lr}
\toprule
Coppon type & Cost\\\midrule
5\% discount&5\\
5\% cashback&5\\
10\% discount&10\\
10\% cashback&10\\
15\% discount&15\\
300 yen discount&13\\
\bottomrule
\end{tabular}
\end{table}

\subsection{Coupon allocation strategies}

We compare the performance of the following coupon allocation strategies:
\begin{description}
\item[Random:] allocating coupons to customers randomly until the budget is used up; 
%\color{blue}
\item [MCK model:] the multiple-choice knapsack model; 
%\color{black}
\item [MVO model:] the mean--variance optimization model;  
\item [RO model:] the robust optimization model.
\end{description}
The MCK, MVO, and RO problems were solved using the mathematical optimization solver Gurobi Optimizer Ver.~11.0.1 (\url{https://www.gurobi.com/}).    %SCS~\cite{o2016conic}, a numerical optimization package in the CVXPY modeling language. 
For the MVO and RO models, we computed an optimal solution $\bm{w}$ and allocated coupon $j \in J$ to $\lfloor w_{kj} \rfloor$ customers, who were extracted randomly from cluster $k \in K$. 
%\textcolor{blue}{Note that we dispense with this random extraction phase for Knapsack model since the variables are associated with not clusters but customers.}

The following hyperparameters in the MVO and RO models were tuned through 5-fold cross-validation using the training dataset:
\begin{align*}
%& \mbox{number of clusters:}~|K| \in \{10,100 \}; \\
& \lambda \in \{0,0.001,0.002,\ldots,0.010 \}~\mbox{(risk aversion)};\\
& \alpha \in \{0, 0.2,0,4,0.6,0.8, 1.0\}~\mbox{(interval scale)};\\
& \Gamma=\beta |K \times J|,~\beta \in \{0, 0.2,0.4,0.6, 0.8\}~\mbox{(cardinality)}. 
\end{align*}
The best configurations of these hyperparametes were $\lambda = 0$ for $|K| \in \{10,100\}$ in the MVO model; and $(\alpha, \beta) = (0.5, 0.8)$ for $|K|=10$, and $(\alpha, \beta) = (1.0, 0.8)$ for $|K|=100$ in the RO model. 

The unit cost ($c_j$) of each coupon $j \in J$ was set according to the discount and cashback rates, as given in Table~\ref{tbl:coupon}. 
The total budget ($B$) for coupon allocation was chosen from 11 values ranging from 400,000 to 1,600,000. 

%\color{blue}
For practical purposes, we add the following lower/upper bound constraints on the allocated proportion of each coupon $j \in J$:
\begin{equation*}
%L \sum_{j \in J} \sum_{k \in K} w_{kj} \le 
%\sum_{k \in K} w_{kj}
%\le U \sum_{j \in J} \sum_{k \in K} w_{kj} \quad (k \in K). 
L \le 
\frac{1}{|I|} \sum_{k \in K} w_{kj}
\le U \quad (j \in J),  
\end{equation*}
where $L$ and $U$ are lower and upper bounds of the allocated coupon proportions. 
These constraints are required (i) to ensure fairness among coupons targeting different products/services, or (ii) to collect an adequate size of sample to estimate each coupon effect. 
We test the two situations without the lower/upper bounds and with the lower/upper bounds $(L,U) = (0.05,0.50)$. 
\color{black}

\begin{table*}
\centering
\caption{Results of customer segmentation ($|K|=10$)}
\label{tbl:cluster}
%\small
\begin{tabular}{rrrrrrrr}\toprule		
&&\multicolumn{6}{c}{Average uplift value~~($\pm$ Standard error)}\\\cmidrule{3-8}
No.	&	Size	&	5\% discount	&	5\% cashback	&	10\% discount	&	10\% cashback	&	15\% discount	&	300 yen discount	\\	\midrule
1&67,090&0.52\ \ \ ($\pm$0.01)&0.19\ \ \ ($\pm$0.01)&1.27\ \ \ ($\pm$0.02)&1.27\ \ \ ($\pm$0.02)&1.32\ \ \ ($\pm$0.02)&1.26\ \ \ ($\pm$0.02)\\
2&49,621&0.56\ \ \ ($\pm$0.01)&0.21\ \ \ ($\pm$0.01)&1.34\ \ \ ($\pm$0.02)&1.32\ \ \ ($\pm$0.02)&1.38\ \ \ ($\pm$0.02)&1.33\ \ \ ($\pm$0.02)\\
3&7,745	&0.58\ \ \ ($\pm$0.03)&0.23\ \ \ ($\pm$0.02)&1.35\ \ \ ($\pm$0.05)&1.31\ \ \ ($\pm$0.06)&1.36\ \ \ ($\pm$0.06)&1.33\ \ \ ($\pm$0.05)\\
4&4,348	&0.64\ \ \ ($\pm$0.05)&0.26\ \ \ ($\pm$0.03)&1.45\ \ \ ($\pm$0.06)&1.42\ \ \ ($\pm$0.05)&1.46\ \ \ ($\pm$0.05)&1.44\ \ \ ($\pm$0.06)\\
5&2,059	&0.51\ \ \ ($\pm$0.08)&0.23\ \ \ ($\pm$0.04)&1.23\ \ \ ($\pm$0.11)&1.24\ \ \ ($\pm$0.11)&1.29\ \ \ ($\pm$0.12)&1.18\ \ \ ($\pm$0.11)\\
6&845	&0.77\ \ \ ($\pm$0.08)&0.28\ \ \ ($\pm$0.05)&1.57\ \ \ ($\pm$0.15)&1.57\ \ \ ($\pm$0.15)&1.57\ \ \ ($\pm$0.15)&1.57\ \ \ ($\pm$0.15)\\
7&115	&0.10\ \ \ ($\pm$0.46)&$-$0.26\ \ \ ($\pm$0.46)&0.88\ \ \ ($\pm$0.46)&0.88\ \ \ ($\pm$0.46)&0.89\ \ \ ($\pm$0.46)&0.88\ \ \ ($\pm$0.46)\\
8&100	&0.85\ \ \ ($\pm$0.14)&0.47\ \ \ ($\pm$0.13)&1.66\ \ \ ($\pm$0.15)&1.66\ \ \ ($\pm$0.15)&1.79\ \ \ ($\pm$0.21)&1.66\ \ \ ($\pm$0.15)\\
9&62	&0.65\ \ \ ($\pm$0.06)&0.33\ \ \ ($\pm$0.06)&1.37\ \ \ ($\pm$0.09)&1.37\ \ \ ($\pm$0.09)&1.38\ \ \ ($\pm$0.09)&1.37\ \ \ ($\pm$0.09)\\
10&15	&1.15\ \ \ ($\pm$0.26)&0.82\ \ \ ($\pm$0.26)&1.97\ \ \ ($\pm$0.37)&1.97\ \ \ ($\pm$0.37)&1.94\ \ \ ($\pm$0.36)&1.97\ \ \ ($\pm$0.37)\\	\bottomrule
\end{tabular}																
\end{table*}

\begin{figure*}[tb]
\centering
\begin{subfigure}[t]{0.495\textwidth}
\centering
\includegraphics[width=0.99\textwidth]{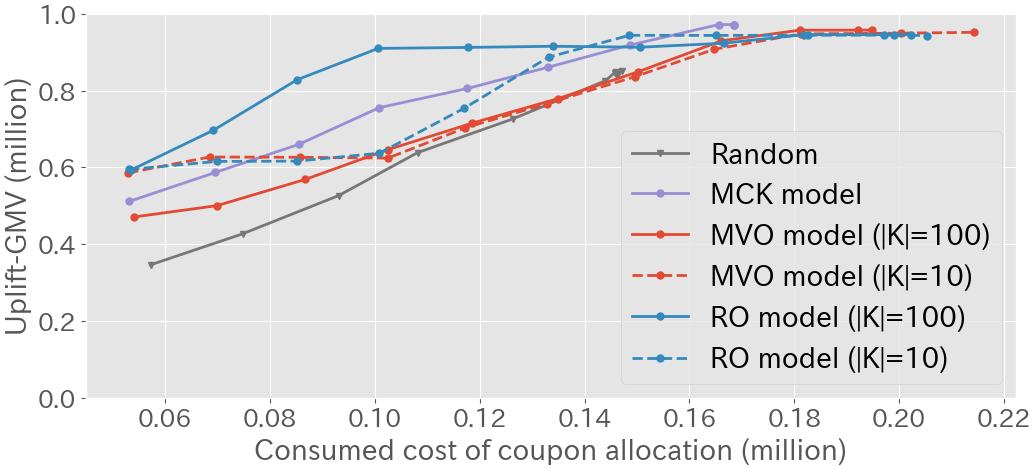}
\caption{Without the lower/upper bounds}\label{fig:proposed_vs_previous_a}
\end{subfigure}
\begin{subfigure}[t]{0.495\textwidth}
\centering
\includegraphics[width=0.99\textwidth]{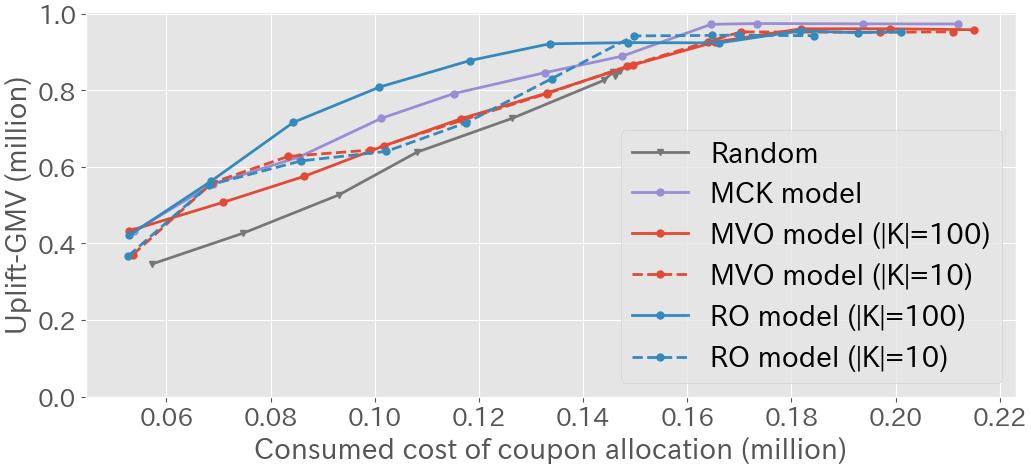}
\caption{With the lower/upper bounds $(L,U)=(0.05,0.50)$}\label{fig:proposed_vs_previous_b}
\end{subfigure}
\caption{Uplift-GMV with respect to the consumed cost}
\label{fig:proposed_vs_previous}
\end{figure*}

\subsection{Evaluation methodology}
We trained machine learning models for coupon effect estimation using the training dataset. 
We next applied the trained models to the testing dataset to estimate coupon effects on customers, and these estimates were used for customer segmentation. 
We solved the optimization models to determine coupon allocation strategies for the testing dataset and then estimated the associated uplift (CATE) of GMV in the following way. 

The repetition of online experiments (e.g., A/B testing), which are expensive and may impair customer experience, is usually difficult. 
We thus use an offline evaluation metric following the concept of the transformed outcome \cite{athey2015machine} based on our dataset of randomly distributed coupons. 
Let $W_{ij} \in \{0,1\}$ be the binary treatment indicator; $W_{ij} = 1$ if the same coupon $j$ is allocated to customer $i$ in the past random distribution and our allocation strategy in the testing dataset, and $W_{ij} = 0$ otherwise. 
The uplift of GMV for the testing dataset by allocating coupon $j$ to customer $i$ is then given by 
\begin{align*}
\label{Uplift_gmv}
% \text{Uplift-GMV}_i = &\text{GMV}_i \frac{W_i - p}{p(1-p)}
\text{Uplift-GMV}_{ij} = \frac{W_{ij} - p}{p(1-p)} \text{GMV}_i,
\end{align*}
where $p$ is the treatment probability in the case of complete randomization; namely, $p = 1/7~(=88{,}000/616{,}000)$ in our case. 
\citet{athey2015machine} proved that the expectation of this uplift value is equal to the CATE. 

\subsection{Results of coupon effect estimation}
We estimated the coupon effect using the S-Learner~\cite{kunzel2019metalearners} with LightGBM~\cite{ke2017lightgbm}, a framework of gradient boosting decision trees.  
The hyperparameters of the LightGBM were tuned through 5-fold cross-validation using Optuna~\cite{akiba2019optuna}, a library for Bayesian optimization. 

Figure~\ref{fig:up} shows uplift curves of GMV for six types of coupons and values of the area under the uplift curve (AUUC) as a common metric of uplift modeling~\cite{devriendt2018literature}. 
We can see from the uplift curves that the cumulative uplift attained by each coupon was larger than the baseline (i.e., random treatment). 
The AUUC of the 10\% cashback (i.e., $\mathrm{AUUC}=499.6$) was the largest, and that of the 300 yen discount (i.e., $\mathrm{AUUC}=137.8$) was the smallest. 
This implies that the high percentage of cashback is most effective, whereas the fixed amount of discounts is not very effective in increasing sales. 

\subsection{Results of customer segmentation}

For customer segmentation, we used a Gaussian mixture model~\cite{bishop2006pattern} implemented by the sklearn.mixture module in the Python scikit-learn library.  
We set $|K| \in \{5, 10, 50, 100\}$ as the number of clusters and employed the estimated uplifts of six types of coupons as customer features for clustering.

Table~\ref{tbl:cluster} gives the cluster size (i.e., the number of customers) and the average uplift value of each coupon for the testing dataset when $K=10$. 
Large clusters (e.g., clusters 1, 2, and 3) had relatively low uplift values for all types of coupons, whereas some small clusters (e.g., clusters 4, 6, 8, and 10) exhibited high uplift values for many types of coupons. 

\begin{figure}[t]
\centering
\begin{subfigure}[t]{0.4\textwidth}
\centering
\includegraphics[width=\textwidth]{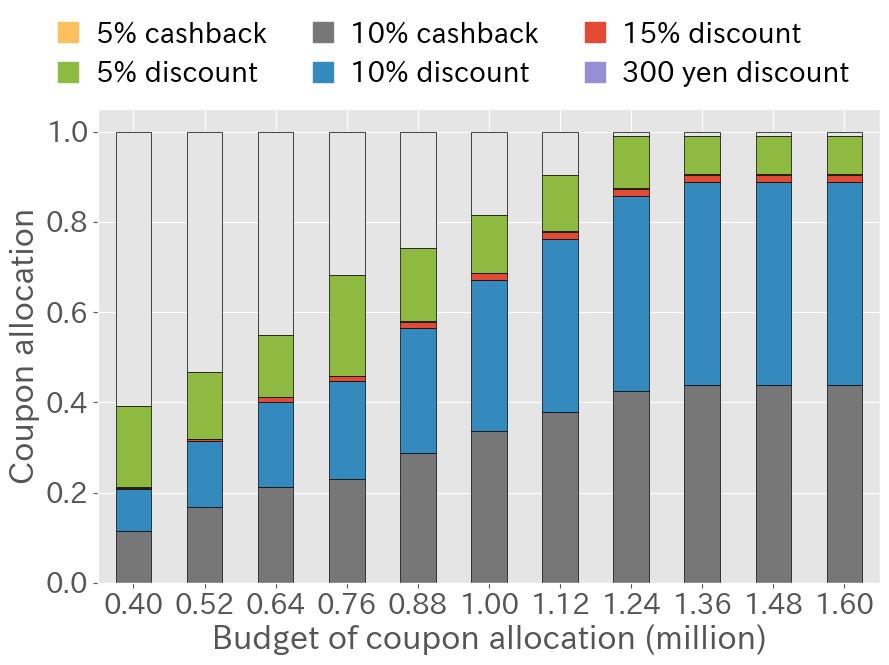}
%{images/previous_allocation.jpg}%
\caption{%\color{blue} 
MCK model}\label{fig:allocation_wo_a}
\end{subfigure}\\[0.3cm]
\begin{subfigure}[t]{0.4\textwidth}
\includegraphics[width=\textwidth]{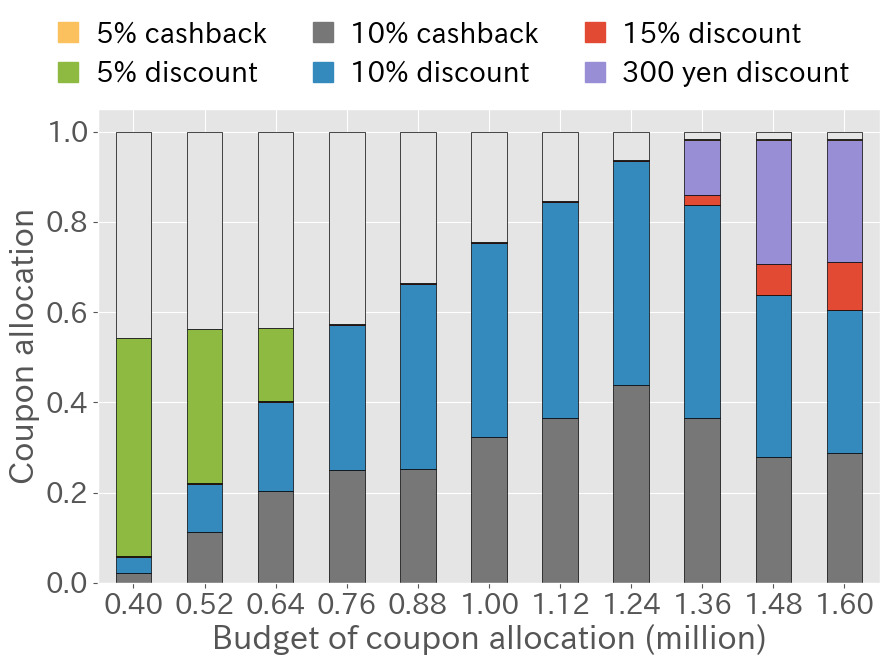}
\caption{MVO model ($|K|=100$)}\label{fig:allocation_wo_b}
\end{subfigure}\\[0.3cm]
\begin{subfigure}[t]{0.4\textwidth}
\includegraphics[width=\textwidth]{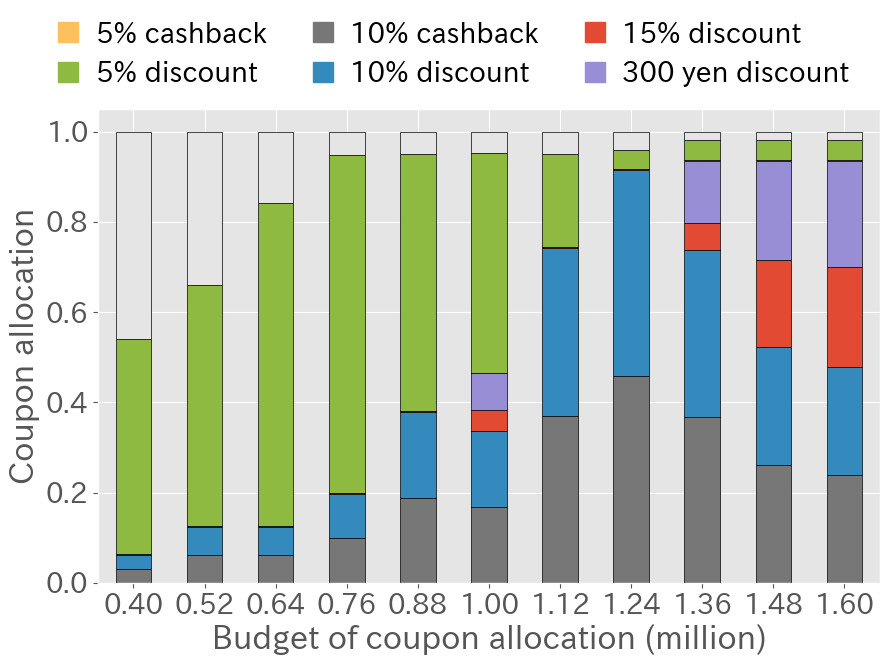}
\caption{RO model ($|K|=100$)}\label{fig:allocation_wo_c}
\end{subfigure}
\caption{Allocated coupons without the lower/upper bounds with respect to the budget}
\label{fig:allocation_wo}
\end{figure}

\begin{figure}[t]
\centering
\begin{subfigure}[t]{0.4\textwidth}
\centering
\includegraphics[width=\textwidth]{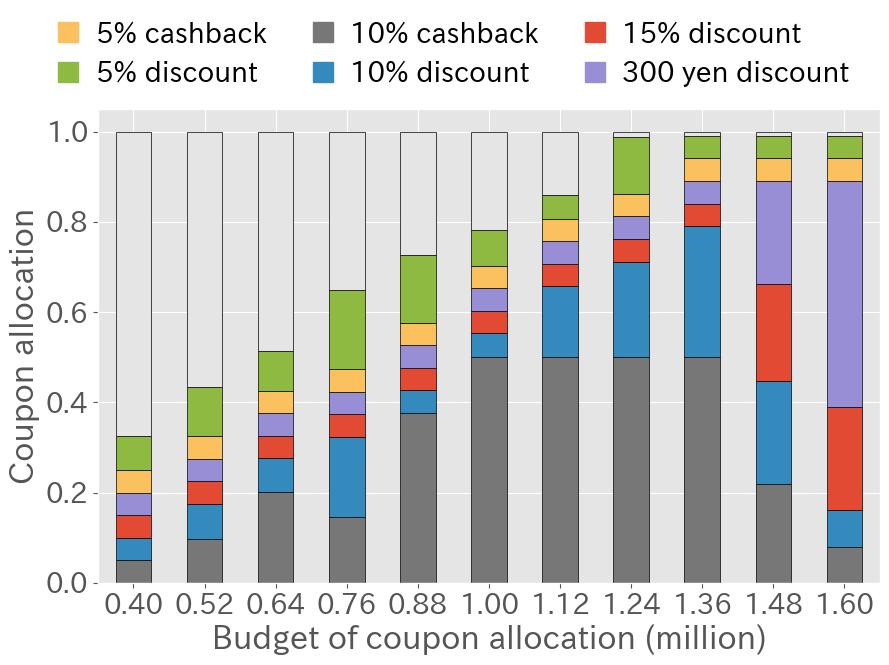}
%{images/newimage2.png}
\caption{%\color{blue} 
MCK model}\label{fig:allocation_w_a}
\end{subfigure}\\[0.3cm]
\begin{subfigure}[t]{0.4\textwidth}
\includegraphics[width=\textwidth]{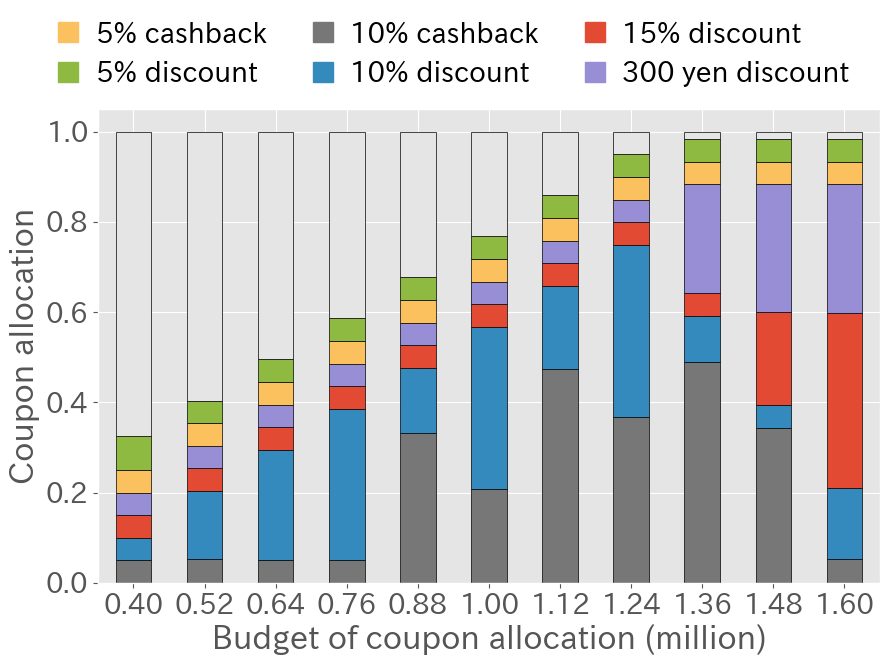}
\caption{%\color{blue} 
MVO model ($|K|=100$)}\label{fig:allocation_w_b}
\end{subfigure}\\[0.3cm]
\begin{subfigure}[t]{0.4\textwidth}
\includegraphics[width=\textwidth]{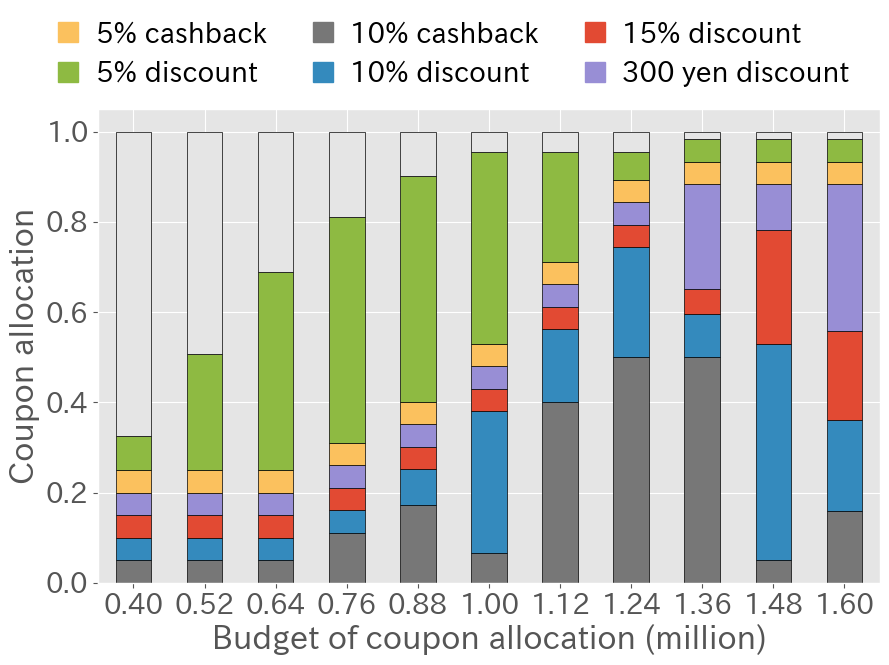}
\caption{%\color{blue} 
RO model ($|K|=100$)}\label{fig:allocation_w_c}
\end{subfigure}
\caption{%\color{blue} 
Allocated coupons with the lower/upper bounds $(L,U) = (0.05, 0.50)$ with respect to the budget}
\label{fig:allocation_w}
\end{figure}

\subsection{Results of coupon allocation}

%\color{blue}
%\susubsection{Change in \text{Uplift-GMV} with respect to budget}
Figure~\ref{fig:proposed_vs_previous} shows the uplift-GMV produced by each coupon allocation strategy with respect to the consumed cost of coupon allocation. 
First, we focus on the results without the lower/upper bounds (Figure~\ref{fig:proposed_vs_previous_a}).
When $|K|=10$, the RO model attained higher uplift-GMV values than did the MVO model for the consumed cost from 100,000 to 160,000. 
When $|K|=100$, the RO model substantially outperformed the MCK and MVO models for the consumed cost from 60,000 to 140,000. 
The RO model consistently outperformed the random coupon allocation, and the difference in the uplift-GMV between these strategies got larger as the consumed cost decreased. 
These results verify that the RO model is very effective in increasing sales especially when the budget of coupon allocation is limited. 

Next, we move on to the results with the lower/upper bounds (Figure~\ref{fig:proposed_vs_previous_b}). 
The RO model with $|K|=100$ still attained the highest uplift-GMV values among the coupon allocation strategies for the consumed cost from 60,000 to 140,000. 
However, the difference in the uplift-GMV among these strategies got smaller in Figure~\ref{fig:proposed_vs_previous_b} than in Figure~\ref{fig:proposed_vs_previous_a} because the lower/upper bounds make allocation strategies less flexible, thereby tending to produce similar results.

Figures~\ref{fig:allocation_wo} and \ref{fig:allocation_w} show the proportions of coupon types allocated by the MCK, MVO and RO models with respect to the budget of coupon allocation. 
First, we focus on the results without the lower/upper bounds (Figure~\ref{fig:allocation_wo}). 
For the budget from 640,000 to 1,000,000, although the MCK model and the MVO model ($|K|=100$) mainly allocated 10\% discount and 10\% cashback coupons to about half of the customers, the RO model ($|K|=100$) mainly allocated 5\% discount coupons to more customers. 
When the budget was over 1,360,000, the MCK model still mainly allocated 10\% discount and 10\% cashback coupons to most customers, and the MVO model ($|K|=100$) and the RO model ($|K|=100$) started to allocate 300 yen discount and 15\% discount coupons. 

\begin{table*}[h!]
\centering
\caption{Coupons allocated by the RO model ($|K|=10$, $B=1{,}000{,}000$) {%\color{blue} 
without the lower/upper bounds}}
\label{tbl:allocation}
\small
\begin{tabular}{rrrrrrrr}\toprule		
&&\multicolumn{6}{c}{Number of allocated coupons}\\\cmidrule{3-8}
No.	&	Size	&	5\% discount	&	5\% cashback	&	10\% discount	&	10\% cashback	&	15\% discount	&	300 yen discount	\\	\midrule
1	&	67,090	&	0      &	0	&	33,544	&	33,544	&	0	&	0	\\%&	2	\\	
2	&	49,621	&	40,619 &	0	&	0	&	0	&	0	&	0	\\%&	9,002	\\	
3	&	7,745	&	0       &	0	&	3,872	&	3,872	&	0	&	0	\\%&	1	\\	
4	&	4,348	&	0       &	0	&	2,174	&	2,174	&	0	&	0	\\%&	0	\\	
5	&	2,059	&	0       &	0	&	0	&	0	&	0	&	0	\\%&	2,059	\\	
6	&	845	    &	845       &	0	&	0	&	0	&	0	&	0	\\%&	0	\\	
7	&	115	    &	0	&	0	&	0	&	0	&	0	&	0	\\%&	115	\\	
8	&	100	    &	0	&	0	&	0	&	0	&	0	&	0	\\%&	100	\\	
9	&	62      &	0	&	0	&	31	&	31	&	0	&	0	\\%&	0	\\	
10	&	15	    &	0	&	0	&	0	&	0	&	15	&	0	\\ \bottomrule\\%&   0	\\	\bottomrule
\end{tabular}
\end{table*}

Next, we move on to the results with the lower/upper bounds (Figure~\ref{fig:allocation_w}). 
In this case, each type of coupons must account for $5\%$ of the allocated coupons due to the lower bound constraints. 
Since there are six types of coupons, which must account for $30\%$ of the allocated coupons, and the problem is thus how to manage the remaining $70\%$ of coupons. 
The MCK model and the MVO model ($|K|=100$) mainly allocated 10\% discount and 10\% cashback coupons.  
For the budget from 640,000 to 1,000,000, in contrast, the RO model ($|K|=100$) allocated 5\% discount coupons to many customers.  
\color{black}

For reference, Table~\ref{tbl:allocation} gives the number of coupons allocated to each customer segment by the RO model ($|K|=10$, $B=1{,}000{,}000$)
%\color{blue}
without the lower/upper bounds. 
%\color{black}
The 5\% discount, 10\% discount, and 10\% cashback coupons were allocated to a majority of customers, and this allocation is marked by the effective use of the 5\% discount coupons. 
This means that the RO model boosted sales by giving low-cost coupons to a broad range of customers. 

\begin{figure}[t]
  \centering
  \begin{subfigure}{0.47\textwidth}
  \includegraphics[width=\textwidth]{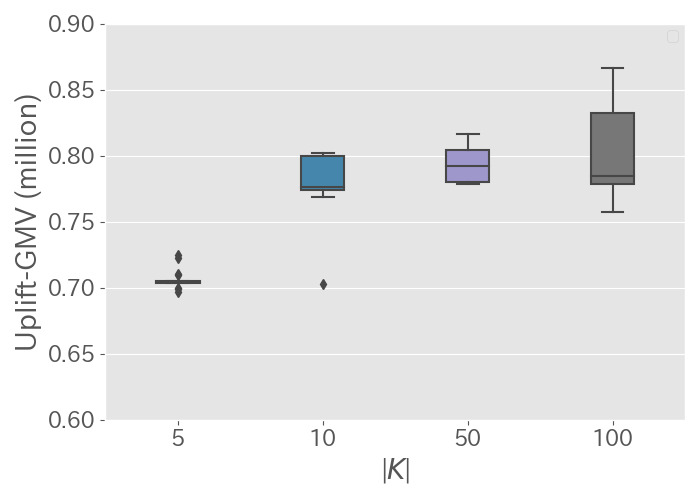}
  \caption{With respect to the number of clusters $|K|$}
  \label{fig:box_k}
  \end{subfigure}\\[0.3cm]
  \begin{subfigure}{0.47\textwidth}
  \includegraphics[width=\textwidth]{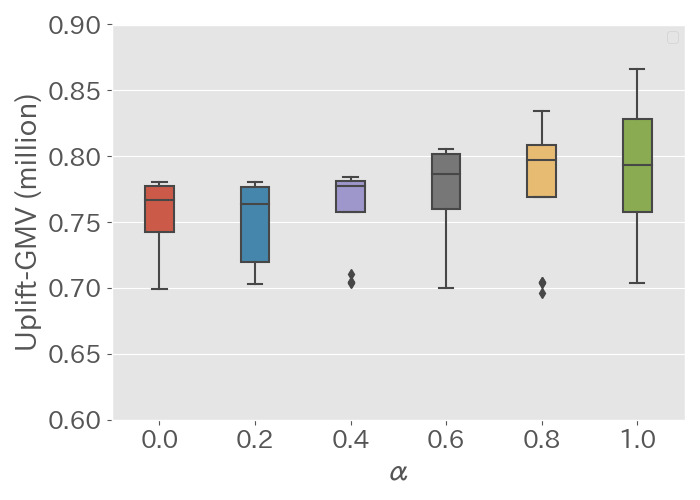}
  \caption{With respect to the interval scale $\alpha$}
  \label{fig:box_alpha}
  \end{subfigure}\\[0.3cm]
  \begin{subfigure}{0.47\textwidth}
  \includegraphics[width=\textwidth]{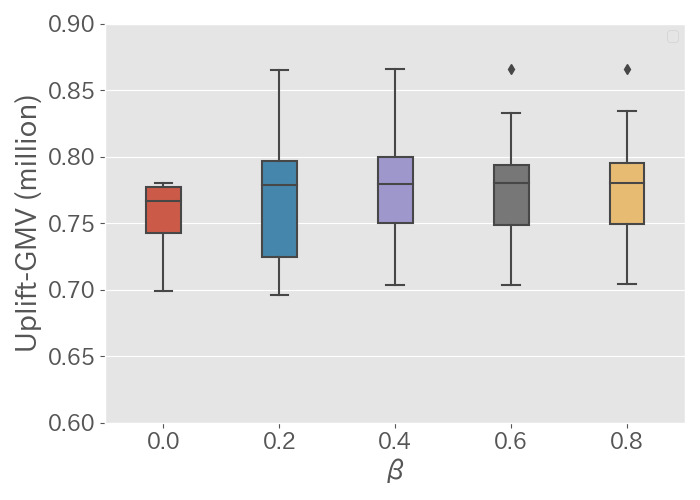}
  \caption{With respect to the cardinality $\Gamma = \beta |K \times J|$}
  \label{fig:box_beta}
    \end{subfigure}
    \caption{Uplift-GMV of the RO model %{\color{blue} 
    without the lower/upper bounds %} 
    with respect to the three hyperparameters}
    \label{fig:box}
\end{figure}

Next, we examine the uplift performance of the RO model 
%\color{blue}
without the lower/upper bounds 
%\color{black}
for various hyperparameter values. 
Figure~\ref{fig:box} shows boxplots of the uplift-GMV with respect to three hyperparameters.   
Note here that each uplift-GMV value was averaged over the 11 values of the total budget ($B$) and corresponded to a different configuration of the hyperparameters $(K,\alpha,\beta)$. 

Figure~\ref{fig:box_k} shows that high uplift-GMV can be achieved by setting $|K| \ge 10$ as the number of clusters.
This implies that $|K|=5$ was too small to make effective coupon allocation based on customer segmentation because a larger number of clusters enable more personalized coupon allocation.  
Figure~\ref{fig:box_alpha} shows that the uplift-GMV increased with the interval scale of the uncertainty set. 
It is clear from this figure that coupon allocation can be improved via robust optimization; this is probably because accurate estimation of coupon effects is very difficult, and therefore, coupon allocation can be improved by considering the uncertainty of estimated coupon effects via robust optimization.  
Figure~\ref{fig:box_beta} shows that the uplift-GMV can be increased by setting $\beta > 0$ for the cardinality $\Gamma = \beta |K \times J|$ in the uncertainty set. 
In contrast, the uplift-GMV was not sensitive to $\Gamma = \beta |K \times J|$ when $\beta \ge 0.2$. 

\section{Conclusion}
\label{conclusion}

We applied a robust portfolio optimization model to the coupon allocation problem, which aims at optimally allocating coupons to customers subject to a budget constraint on an e-commerce website. 
This framework is highly scalable because the coupon allocation (i.e., large-sized combinatorial optimization) problem can be reduced to a small-sized continuous portfolio optimization problem.  
Our numerical results demonstrated that the robust optimization model substantially outperformed 
%\color{blue}
the multiple-choice knapsack model, 
%\color{black}
the mean--variance optimization model, and random coupon allocation. 

To the best of our knowledge, no prior studies have considered a robust portfolio optimization model for allocating various types of coupons. 
We are also the first to verify that uplift of sales can be achieved by robust optimization through numerical experiments using actual data of randomly distributed coupons. 
Our study highlights the practical effectiveness of the robust portfolio optimization in the coupon allocation problem. 

A future direction of study will be to consider other types of robust optimization models~\cite{bertsimas2011theory,Rahimian_2022} for coupon allocation. 
Another direction of future research will be to extend our optimization model to dynamic coupon allocation (e.g., using the reinforcement learning~\cite{liu2023dynamic}). 

\paragraph{Data Availability Statement}
Due to the nature of this research, participants of this study did not agree for their data to be shared publicly, so supporting data is not available.

\paragraph{Conflict of Interest}
The authors have no conflicts of interest to declare.

\bibliography{aaai23}

\end{document}